\documentclass[twocolumn ,pra,longbibliography,a4paper,superscriptaddress,tightenlines,noeprint,10pt]{revtex4-2}
\RequirePackage{silence}
\RequirePackage{float}
\setcitestyle{compressed}
\usepackage[utf8]{inputenc}
\usepackage[USenglish]{babel}
\usepackage[T1]{fontenc}
\usepackage{lmodern}
\usepackage{array,multirow}
\usepackage{algorithm}
\usepackage{algorithmic}
\usepackage{graphicx,epsf}
\usepackage[caption=false]{subfig}
\usepackage[toc,page,header]{appendix}
\usepackage{minitoc}
\usepackage{tikz}
\usetikzlibrary{quantikz}
\usepackage[bookmarksnumbered,hypertexnames=false,colorlinks,colorlinks=true,linkcolor=blue,urlcolor=black,citecolor=blue,anchorcolor=green]{hyperref}
\usepackage{bbm,braket,microtype,mathrsfs,amsmath,amssymb,color,amsthm,mathtools,fullpage,enumitem,ae,aecompl,bm,thm-restate}
\usepackage{tikz}
\usepackage{lipsum}
\usepackage[justification=justified]{caption}
\usepackage{color}
\usepackage[capitalize]{cleveref}
\usepackage{silence}
\usepackage{newpxtext}
\usepackage{mathtools}

\allowdisplaybreaks
\newcommand{\be}{\begin{equation}}
\newcommand{\ee}{\end{equation}}
\newcommand{\ba}{\begin{eqnarray}}
\newcommand{\ea}{\end{eqnarray}}
\DeclareMathOperator{\tr}{tr}
\DeclareMathOperator{\card}{card}

\newcommand{\ignore}[1]{}
\newcommand{\st}[1]{\ket{#1}\bra{#1}}

\newcommand{\aver}[1]{ \left\langle  {#1}  \right\rangle }

\def\CC{{\rm\kern.24em \vrule width.04em height1.46ex depth-.07ex
   \kern-.29em C}}
\def\P{{\rm I\kern-.25em P}}
\def\RR{{\rm
        \vrule width.04em height1.58ex depth-.0ex
        \kern-.04em R}}

\def\bbbc{{\mathchoice {\setbox0=\hbox{$\displaystyle\rm C$}\hbox{\hbox
to0pt{\kern0.4\wd0\vrule height0.9\ht0\hss}\box0}}
{\setbox0=\hbox{$\textstyle\rm C$}\hbox{\hbox
to0pt{\kern0.4\wd0\vrule height0.9\ht0\hss}\box0}}
{\setbox0=\hbox{$\scriptstyle\rm C$}\hbox{\hbox
to0pt{\kern0.4\wd0\vrule height0.9\ht0\hss}\box0}}
{\setbox0=\hbox{$\scriptscriptstyle\rm C$}\hbox{\hbox
to0pt{\kern0.4\wd0\vrule height0.9\ht0\hss}\box0}}}}
\def\bbbz{{\mathchoice {\hbox{$\sf\textstyle Z\kern-0.4em Z$}}
{\hbox{$\sf\textstyle Z\kern-0.4em Z$}}
{\hbox{$\sf\scriptstyle Z\kern-0.3em Z$}}
{\hbox{$\sf\scriptscriptstyle Z\kern-0.2em Z$}}}}

\newlength{\fighskip} \fighskip=2pt
\newlength{\figvskip} \figvskip=1pt

\usepackage{hyperref}

\begin{document}
\setcounter{secnumdepth}{3}
\title{Magic-state resource theory for the ground state of the transverse-field Ising model}
\author{Salvatore F.E. Oliviero}\email{s.oliviero001@umb.edu}
\affiliation{Physics Department,  University of Massachusetts Boston,  02125, USA}
\author{Lorenzo Leone}
\affiliation{Physics Department,  University of Massachusetts Boston,  02125, USA}
\author{Alioscia Hamma}
\affiliation{Dipartimento di Fisica Ettore Pancini, Universit\`a degli Studi di Napoli Federico II,
Via Cinthia, I-80126 Napoli, Italy}
\affiliation{Physics Department,  University of Massachusetts Boston,  02125, USA}

\begin{abstract}
Ground states of quantum many-body systems are both entangled and possess a kind of quantum complexity as their preparation requires universal resources that go beyond the Clifford group and stabilizer states. These resources - sometimes described as {\em magic} -  are also the crucial ingredient for quantum advantage. We study the behavior of the stabilizer R\'enyi entropy in the integrable transverse field Ising spin chain. 
We show that the locality of interactions results in a localized stabilizer R\'enyi entropy in the gapped phase thus making this quantity computable in terms of local quantities in the gapped phase, while measurements involving $L$ spins are necessary at the critical point to obtain an error scaling with $O(L^{-1})$. 
\end{abstract}
\maketitle

{\em Introduction.---} Quantum mechanics is different from classical physics in two ways: First, composite quantum systems can exhibit correlations stronger than any classical correlation, i.e. entanglement. Second, because quantum states and operations constitute the bedrock for computation that goes beyond the classical Turing machine model and can outperform classical algorithms\cite{Somma2008ann,Kimble2008internet,Cirac2012goals,Bravy2018adv,acin2018tech,Arute2019sup}. The resource useful for such a quantum advantage consists of those states and operations that go beyond the stabilizer formalism and the Clifford group\cite{Veitch2014resource,Ahmadi2018magic,Wang2019magic,seddon2019magic,Sarkar2020characterization,liu2020many,White2021cft,qassim2021improved,Koukoulekidis2021constraints,Hahn202magic,Saxena2022stab,sewell2022mana}. 
 
 Entanglement has been widely studied in the context of quantum many-body systems\cite{horodecki2009entanglement} from its role in quantum phase transitions\cite{Vidal2003ent,Amico2008Ent,Osterloh2002ent,osborne2002ent,LeHur2007ent}, to issues of simulability\cite{cirac2006matrix,perezgarcia2007matrix,Schuch2008entropy,Wolf2008area,Eisert2006Ent,schuch2010peps,Hastings2007area,chen2012groundstates,harrow2017local}, to the onset or thermalization and chaos in closed quantum systems\cite{srednicki1994chaos,popescu2006thermal,popescu2006entanglement,rigol2008thermalization,santos2010onset,yang2017entanglement,Neill2016ergodic,rigol2016thermalization}, the structure of exotic quantum phases of matter\cite{Hamma2005ground,Kitaev2006topological, Levin2006topological,Chung2010topo,Mezzacapo12ground,Fradkin2007topo,Zhang2011Topo,Kim2012tee,Jiang2012heisenberg,Jiang2012topological,Furukawa2007topo,Hamma2008topo,Dusuel2011robust,Jadamagni2018robust,Cincio2013Topo,oliviero2022topo}, and black hole dynamics\cite{lloyd1988black,srednicki1994chaos,rigol2016thermalization}. On the other hand, magic state resource theory has only very recently been the object of investigation in the field of quantum systems with many particles\cite{liu2018entanglement,liu2018generalized}. This is mainly due to the difficulty of computing non-stabilizerness for high-dimensional spaces\cite{goto2021chaos}. Recently, though, the authors of this paper have proposed the stabilizer R\'enyi entropy as a more amenable way of computing non-stabilizerness based on the R\'enyi entropy associated to the decomposition of a state in the Pauli basis\cite{leone2021renyi}, which has also led to its experimental measurement\cite{oliviero2022measuring,haug2022magic,leone2022magic}. 
 
 In this paper, we set out to show the role that magic state resource theory plays in the ground state of local integrable quantum many-body systems. The model studied here is the transverse field Ising model for a spin one-half chain with $N$ sites. We show how to compute the stabilizer R\'enyi entropy in terms of the ground-state correlation functions. In this way, we see how the decay of correlation functions influences the many-body non-stabilizerness. Away from the critical point, where the ground state is weakly entangled and two-point correlation functions decay exponentially, it is possible to estimate the stabilizer R\'enyi entropy by single spin measurements reliably. At the critical point, on the other hand, one needs to measure an entire block of spins to obtain a reliable estimate, with an error scaling with a characteristic power-law $O(L^{-1})$. The result is of notable importance for experimental measurements of non-stabilizerness in a quantum many-body system, as in a gapped phase this can be performed by few spin measurements (even just a single spin).
 
As a last comment, our findings can be relevant for the investigation of the emergence of quantum spacetime in the context of AdS+CFT correspondence: in a recent paper\cite{goto2021chaos}, the authors speculate on the role of non-stabilizerness in AdS+CFT, and argue that it is a key ingredient to fill the complex structure of the AdS black hole interior, dual to a CFT state. Magic state resource theory indeed reveals itself as an important piece of information that cannot be detected by only looking at the entanglement. In this context, it is well known that a quantum many-body system at the criticality is described by a CFT\cite{Vidal2003ent,Calabrese2004ent}. Our results thus give insights regarding the role played by non-stabilizerness in AdS+CFT correspondence: This resource is delocalized in spatial degrees of freedom as, at criticality only, it can be extracted by a system containing $L$ spins with an error decaying only polynomially in $L$. From this result, it can be reasonably argued that delocalization of non-stabilizerness is a universal property in CFT quantum states -- being the correlation functions decaying polynomially -- thus revealing fascinating perspectives in the AdS+CFT correspondence.

{\em Setup and model.---} Let us start by briefly reviewing the stabilizer R\'enyi entropy\cite{leone2021renyi}. Consider an $N$-qubit system and the decomposition of a state $\rho$ in the Pauli basis given by
 $\rho=\frac{1}{2^N}\sum_{P\in\mathbb{P}(N)}\tr(P\rho)P$ with $\mathbb{P}(N)$ being the Pauli group. The $2-$stabilizer R\'enyi entropy $M_{2}(\rho)$ is then defined as:
\be
M_{2}(\rho):=-\log_2\mathbb{E}_{\mathcal{P}}\left[\tr^{2}(P\rho)\right]
\label{magicdefinition}
\ee
i.e., as the average of $\tr^2(P\rho)$ on a state-dependent probability distribution defined as $\mathcal{P}(\rho):=\{2^{-N}\tr^{2}(P\rho)\tr^{-1}(\rho^2)\}$. It is interesting to note that for $\rho$  pure, $M_{2}(\rho)$ reduces to the two-R\'enyi entropy of the classical probability distribution $\mathcal{P}(\rho)$ (modulo an offset of $-N$).


We study the behavior of $M_{2}$ in the ground state of the transverse field Ising model for a spin one-half $N$-site chain with
Hamiltonian
\be
H(\lambda)=-\sum_{i=1}^{N}(\sigma_{i}^{x}\sigma_{i+1}^{x}+\lambda\sigma_{i}^{z})
\label{isinghamiltonian}
\ee
where $\sigma_{i}^{\alpha}$, for $\alpha=x,y,z$, are Pauli matrices defined on the $i$-th site. 
The model displays a quantum phase transition at $\lambda=1$ between a disordered and a symmetry-breaking phase. The critical point corresponds to a conformal field theory with $c=\frac{1}{2}$\cite{sachdev2007handbook}. For $\lambda\rightarrow\infty$ and $\lambda=0$ the Hamiltonian reduces to a stabilizer Hamiltonian\cite{temme2015fast} with stabilizer groups $\mathbb{Z}\rhd\mathbb{P}$ and $\mathbb{X}\rhd\mathbb{P}$ respectively. The model $H(\lambda)$ is  integrable through standard techniques\cite{lieb1961chain, barouch1971xy}. First by a Jordan-Wigner transformation introducing fermionic modes and subsequently by a Fourier and a Bogoliubov transformations\cite{Suzuki2013}. Following these techniques, let us introduce the Majorana operators $A_{l}$ and $B_{l}$:
\be
A_{l}:=\bigotimes_{i<l}\sigma_{i}^{z}\otimes \sigma_{l}^{x};\quad B_{l}:=\bigotimes_{i<l}\sigma_{i}^{z}\otimes \sigma_{l}^{y}.
\ee
These operators  obey the  anti-commutation relations $\{A_{l},A_{l^\prime}\}=\{B_{l},B_{l^\prime}\}=2\delta_{ll^\prime}$ and $\{A_{l},B_{l^\prime}\}=0$. 

The computation of  $M_{2}$ for the ground state $\ket{G(\lambda)}$ of such a class of Hamiltonians relies on the fact that the ground state can be fully characterized by just the two-point correlation functions, by virtue of the Wick theorem: One can compute all the correlation functions of an arbitrary product of Majorana fermions by just knowing the $2-$point correlation functions $\aver{A_{l}A_{l^\prime}}=\aver{B_{l}B_{l^\prime}}=\delta_{ll^\prime}$ and\cite{Suzuki2013} $\aver{A_{l}B_{l^\prime}}\equiv\aver{A_{l}B_{l+r}}\equiv G_r(\lambda)$, where:
\be
G_{r}(\lambda)=-\frac{i}{\pi}\int_{0}^{\pi}\frac{\sin\theta\sin\theta r-(\lambda-\cos\theta)\cos\theta r}{\sqrt{\sin^{2}\theta+(\lambda-\cos\theta)^2}}
\label{2corrfunctions}.
\ee
Indeed, let $\mathcal{C}(\{i\}_{k},\{j\}_{l}):=\aver{A_{i_1}\cdots A_{i_k}B_{j_1}\cdots B_{j_l}}$ be the expectation value on the ground state $\ket{G(\lambda)}$ of an arbitrary ordered product of Majorana fermions, where $\{i\}_{k}:=\{i_{1},\ldots,i_{k}\,|\, N\ge i_{1}>\ldots>i_{k}\ge 1\}$ is a set of ordered indexes ranging over all the sites. The computation of $C(\{i\}_{k},\{j\}_{l})$ can be done through the Pfaffian technique\cite{Caianiello1952} which leads to $\mathcal{C}(\{i\}_{k},\{j\}_{l})=0$ unless $k=l$ and:
\be
\mathcal{C}(\{i\}_{k},\{j\}_{k})=\begin{vmatrix}\aver{A_{i_1}B_{j_1}}& \aver{A_{i_1}B_{j_2}} &\cdots & \aver{A_{i_1}B_{j_k}}\\
\aver{A_{i_2}B_{j_1}} & \aver{A_{i_2}B_{j_2}} &\cdots & \aver{A_{i_2}B_{j_k}}\\
\vdots& \vdots & \ddots & \vdots\\
\aver{A_{i_k}B_{j_1}} & \aver{A_{i_k}B_{j_2}} &\cdots & \aver{A_{i_k}B_{j_k}}
\end{vmatrix}
\label{toeplizmatrix}
\ee
i.e. to compute the generic $2k$-point correlators of Majorana fermions, it is sufficient to compute the determinant of a $k\times k$ matrix, which can be efficiently done numerically by a $poly(k)$ algorithm.

All the $2k$-point correlations functions, can be also obtained by considering the maximum rank $2N$-point correlation function of Majorana fermions $C(\{i\}_N,\{j\}_N)=\aver{A_{1}A_{2}\cdots A_{N}B_{1}B_{2}\cdots B_{N}}$; indeed, it is easy to see that one can obtain any correlation function of order $2k$ by considering any minor of $C(\{i\}_N,\{j\}_N)$ of lower rank $k$. Since a $N\times N$ matrix contain $\binom{N}{k}^{2}$ minors of order $k$, there are $\sum_{k=0}^{N}\binom{N}{k}^{2}=\binom{2N}{N}\simeq  \frac{4^N}{\sqrt{N}}$ nonzero correlation functions of Majorana fermions.

{\em Ground state non-stabilizerness.---}
 In this section, we compute $M_{2}$ in the ground state $\ket{G(\lambda)}$ and discuss some of its properties. To this end, we need the knowledge of all the $4^N$ expectation values of Pauli strings $P\in\mathbb{P}(N)$ on the ground state $\ket{G(\lambda)}$. Except for $\lambda=0, \lambda\rightarrow\infty$, all the other points feature a non-trivial value for the stabilizer R\'enyi entropy because the state cannot be factorized.

It is easy to see that any $P\in\mathcal{P}(N)$ can be written (up to a global phase) as an ordered product of Majorana fermions, as $P\propto A_{i_1}\cdots A_{i_k}B_{j_1}\cdots B_{j_l}$ for some $\{i\}_{k},\{j\}_{l}$, which means that we can write the two-stabilizer R\'enyi entropy for $\ket{G(\lambda)}$ as:
\be
M_{2}(\lambda):=M_{2}(\ket{G(\lambda)})-\log_2\frac{1}{2^N}\sum_{\{i\}_{k},\{j\}_k\le N}C(\{i\}_k,\{j\}_k)^{4}.
\label{magicintermsofcorrelations}
\ee
As the above formula shows, the computation of the non-stabilizerness requires $\sim 4^N$ determinants, which makes the computation exponentially hard in $N$. Let us provide an upper bound to the two-stabilizer entropy given by the zero-stabilizer entropy $M_{0}(\lambda)\ge M_{2}(\lambda)$\cite{leone2021renyi}, which essentially counts the number of nonzero entries $\operatorname{card}(\ket{\psi})$ in the probability distribution $\mathcal{P}(\st{\psi})$ as $M_{0}(\ket{\psi}):=\log_2\card(\ket{\psi})-N$. As explained above, there are $\binom{2N}{N}$ nonzero Majorana correlations functions and thus we can upper bound the two-stabilizer R\'enyi entropy as $M_{2}(\lambda)\lesssim  N-\frac{1}{2}\log_2 N$.

\begin{figure}[ht]
    \centering
    \includegraphics[width=\linewidth]{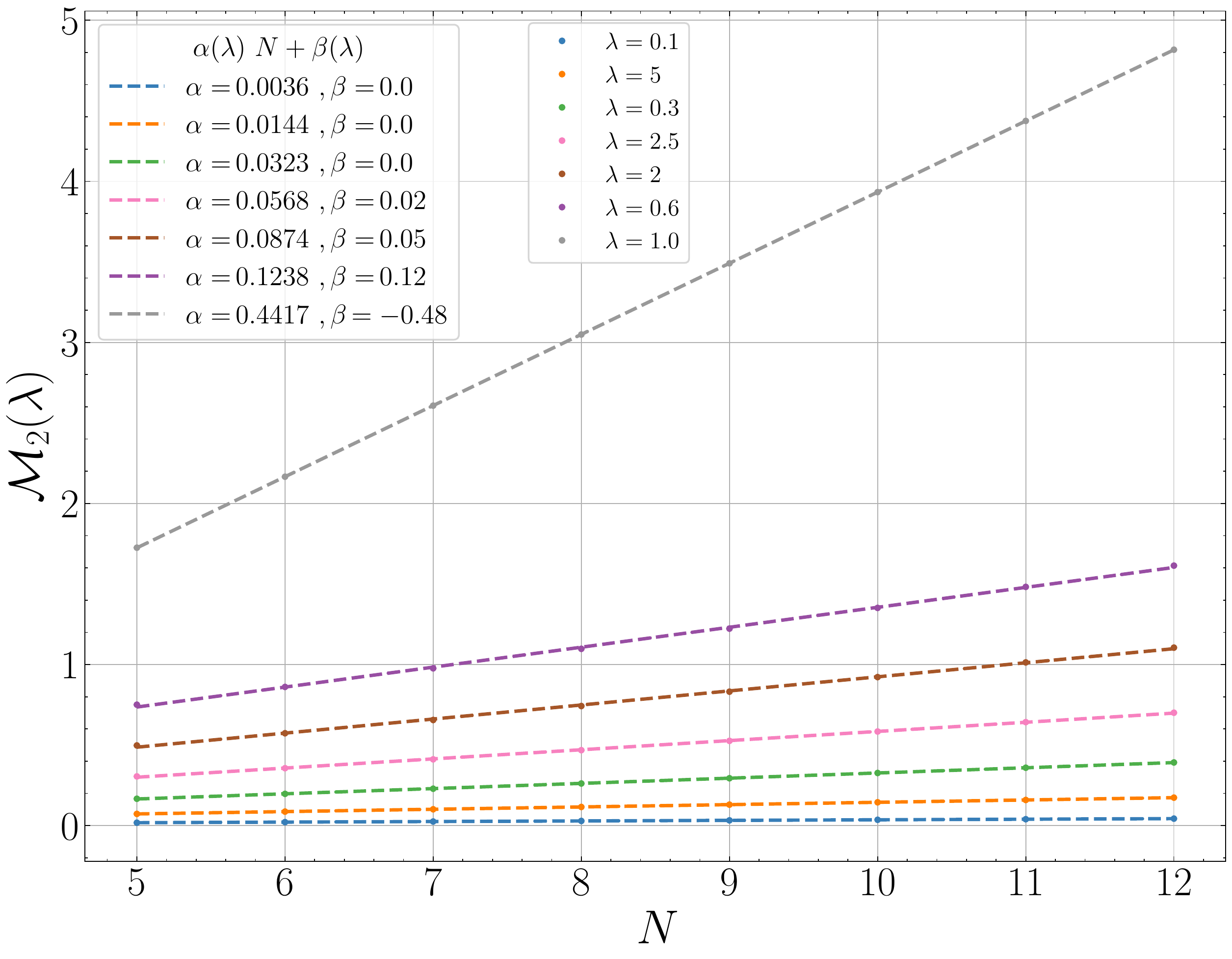}
    \caption{\raggedright Numerical simulations of the stabilizer R\'enyi entropy of the ground state $\ket{G(\lambda)}$ of the Hamiltonian in Eq.~\eqref{isinghamiltonian} for $\lambda=0.1,0.3,0.6$, $\lambda=1$ and $\lambda=2,2.5,5$ as a function of the length of the chain $N\in [5,12]$. The curves are fitted to be straight lines for any $\lambda$s, with slopes $\alpha(\lambda)$ and intercepts $\beta(\lambda)$ fitted in the top-left corner.}
    \label{fig1a}
\end{figure}

We evaluate numerically formula Eq.~\eqref{magicintermsofcorrelations} for $N=5,\ldots, 12$, see Fig.~\ref{fig1a}.
The calculations clearly show a linear behavior of the stabilizer R\'enyi entropy for any $\lambda\neq 0,\infty$:
\be
M_2(\lambda)= \alpha(\lambda)N+\beta(\lambda)
\label{resultfit}
\ee
with both slope $\alpha(\lambda)$ and intercept $\beta(\lambda)$ depending on intensity $\lambda$ of the external magnetic field. In particular, we observe an increasing slope $\alpha(\lambda)$ from $\lambda=0$ towards the criticality at $\lambda=1$, where $\alpha(\lambda)$ approaches its maximum $\alpha(1)\approx0.44$, and then it starts decreasing again in the disordered phase, $\lambda>1$. We thus find agreement with the result in Ref.~\cite{White2021cft}: the ground state at the critical point, and the corresponding $\frac{1}{2}$ CFT, achieves the highest value of non-stabilizerness among the $\lambda$s. However, this result does not tell us the full story, as the behavior of non-stabilizerness with $\lambda$ is quite smooth and is $O(N)$ for every value of $\lambda$. As we show in the following section, the locality of the interactions together with a gap implies that non-stabilizerness is localized, whereas at the critical point non-stabilizerness cannot be resolved by local measurements.

{\em Access non-stabilizerness by local measurements.---} Although more amenable than a minimization procedure\cite{howard2017application}, computing the stabilizer entropy is an exponentially difficult task. However, the locality of the interactions in the Hamiltonian and the presence of a gap results in a fast decay of correlation functions in the ground state, while a power-law characterizes the critical point. One thus wonders if one can exploit this locality to access the stabilizer R\'enyi entropy by local quantities. This results both in the possibility of a realistic experimental measurement of non-stabilizerness in the ground state of quantum many-body systems and a computational advantage.

\begin{figure}[h]
    \centering
    \includegraphics[width=\linewidth]{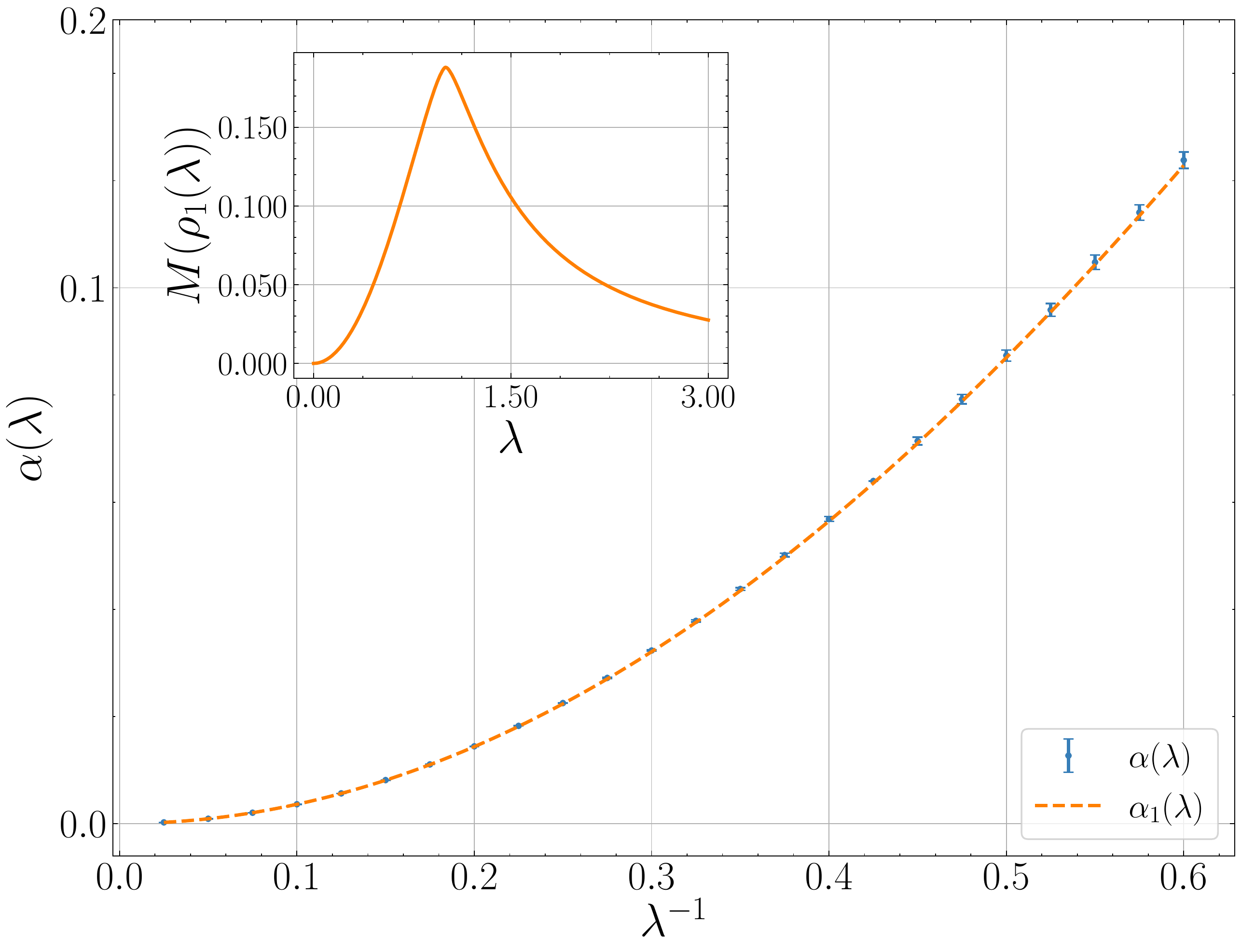}
    \caption{\raggedright Plot of the single spin density of non-stabilizerness $\alpha_{1}(\lambda)$ for $\lambda^{-1}\le 0.6$, computed in Eq.~\eqref{densitymatrix1sitemagic}, versus the density of non-stabilizerness $\alpha(\lambda)$ extracted through the fits in Fig.~\ref{fig1a}.}
    \label{fig1b}
\end{figure}

Let us focus on asymptotic behavior in  $N$, so that  
$M_2(\lambda)\approx \alpha(\lambda)N$. 
We refer to $\alpha(\lambda)$ as the \textit{density of non-stabilizerness}. In the above, $\approx$ stands for 'up to an order $N^{-1}$'. Now, it is clear that if one is able to measure the density $\alpha(\lambda)$, then one accesses the non-stabilizerness of the ground state. Can we measure the density of non-stabilizerness $\alpha(\lambda)$, by just looking at the local properties of the reduced density matrix of $L$ spins?
To answer the question, we first divide the chain of $N$ sites into $N/L$ sub-chains of $L$ first neighbor sites. Consider the following quantum map $\mathcal{L}(\st{GS(\lambda)}^{\otimes N/L})=\bigotimes_{s=0}^{N/L-1}\rho_{L_i}$, where $\rho_{L_{s}}:=\tr_{N-L_{s}}(\st{GS(\lambda)})$ where $L_{s}=(sL+1,\ldots, (s+1)L)$. To estimate the density of non-stabilizerness $\alpha(\lambda)$ of the ground state we thus measure the density $\alpha_{L}(\lambda)$ present in a subsystem of size $L$.  Thanks to the translational invariance of the Hamiltonian in Eq.~\eqref{isinghamiltonian}, all the reduced density matrices are equal to $\rho_{L} \equiv\tr_{N-L_{0}}(\st{GS(\lambda)})$, and thus the local density of non-stabilizerness $\alpha_{L}(\lambda)$ depends on the number of sites $L$ of the sub-chains and not on their locations. Define the $L-$density of non-stabilizerness as:
\be
\alpha_{L}(\lambda):=\frac{1}{L}M_{2}(\rho_{L})
\ee
where $M_{2}(\rho_L)$ is the Stabilizer R\'enyi entropy of the mixed state $\rho_{L}$ (see Eq.~\eqref{magicdefinition}) which in terms of Majorana correlation functions reads:
\be
M_{2}(\rho_{L})=-\log_2\frac{\sum_{\{i\}_{k},\{j\}_k\le L}\mathcal{C}(\{i\}_k,\{j\}_k)^{4}}{\sum_{\{i\}_{k},\{j\}_k\le L}\mathcal{C}(\{i\}_k,\{j\}_k)^{2}}.
\ee
The latter equation, unlike Eq.~\eqref{magicintermsofcorrelations}, contains only correlation functions on at most $L$ sites, thus it does not involve global measurements, rather it involves just measurements on local observables via the reduced density matrix $\rho_L$, which makes it analytically computable for a reasonable $L$. First note that for $L\rightarrow N$, one has $\alpha_{L}(\lambda)\rightarrow\alpha(\lambda)$. Then, how good is the approximation for a finite $L$, and how does it depend on $\lambda$? 
Let us look at the accuracy of the measurement of the $L-$density of non-stabilizerness by looking at the percent error $\epsilon_{L}(\lambda):=\frac{|\alpha(\lambda)-\alpha_{L}(\lambda)|}{\alpha(\lambda)}$ we make by measuring the density of non-stabilizerness via local measurements. We find that, away from the criticality, i.e. in the regions $\lambda\ll1$ and $\lambda\gg1$, $\epsilon_{\lambda}(L)< 0.001$ for any $L$. We thus conclude that, away from the critical point, one can access the non-stabilizerness of the ground state by just measuring the non-stabilizerness of the density matrix of an $O(1)$  of spins, in fact, even a single qubit density matrix $\rho_{1}$. We show the agreement between the the $1-$density of non-stabilizerness $\alpha_{1}(\lambda)$ and the density of non-stabilizerness $\alpha(\lambda)$ in Fig.~\ref{fig1b} for $\lambda>1$. The region $\lambda<1$ features the same behavior, indicating that the non-stabilizerness does not reveal the symmetry of the ground state.

\begin{figure}[h]
    \centering
    \includegraphics[width=\linewidth]{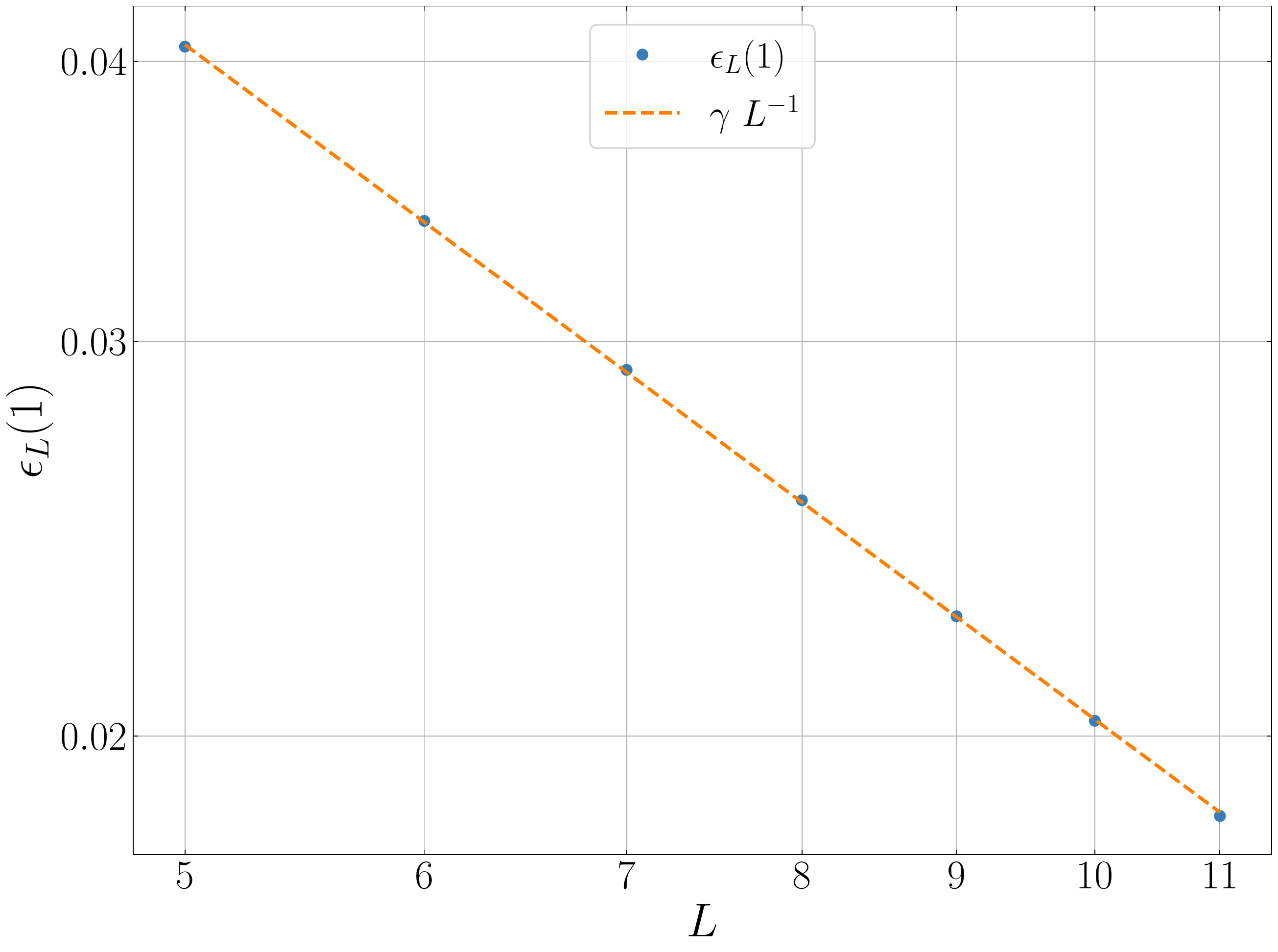}
    \caption{\raggedright Comparison of the error $\epsilon_L (1)$ for $L\in \{5,11\}$ with the fit $\gamma L^{-1}$ with $\gamma=0.2034\pm 0.0002$.}
    \label{fig1c}    
\end{figure}

The situation changes at the critical point, i.e. $\lambda=1$: one finds $\epsilon_{L}(\lambda)=O( L^{-1})$, cfr. Fig.~\ref{fig1c}. The different behaviors of the error, i.e. $O(1)$ vs. $O(L^{-1})$, are reminiscent of different behavior of the entanglement entropy, displaying an area law everywhere, but at the critical point where the entanglement entropy of a density matrix of $L$ spin scales as $\sim\log_2 L$. 

Thus, away from the critical point, the approximation works great also for $L=1$, which can be computed by hand: The single site density matrix reads\cite{osborne2002ent} $\rho_1(\lambda)=\frac{1}{2}(I+\aver{\sigma^{z}}\sigma^{z})$, whose stabilizer R\'enyi entropy:
\be
M(\rho_1(\lambda))=\log_2\frac{1+\aver{\sigma^{z}}^{2}}{1+\aver{\sigma^{z}}^{4}}.
\label{densitymatrix1sitemagic}
\ee
where $|\aver{\sigma^{z}}|=G_{0}(\lambda)$, cfr. Eq.~\eqref{2corrfunctions}; see the inset in Fig.~\ref{fig1b} for a plot.

In the following, we lay down a theoretical argument supporting the fact that measuring the single spin density of non-stabilizerness is already sufficient away from the critical point $\lambda=1$.
It is well known that\cite{Suzuki2013}, away from the criticality (w.l.o.g. let us say $\lambda\gg1$), the two-point correlation functions in Eq.~\eqref{2corrfunctions} decay faster than exponentially with $r$. By making the first order expansion $G_{r}(\lambda)\simeq G_{0}(\lambda)\delta_{r,0}$, one gets a fair approximation of $G_{r}(\lambda)$ as long as the higher terms in $r\neq 0$ are exponentially suppressed. By using the above form of the two-correlation functions to compute higher-order functions as in Eq.~\eqref{toeplizmatrix}, one gets $|\mathcal{C}(\{i\}_{k},\{j_{k}\})|=|G_{0}(\lambda)|^{k}\delta_{\{i\}_k}^{\{j\}_k}$. This means that the only nonzero correlation functions correspond to Pauli operators belonging to the subgroup $\mathbb{Z}\le \mathbb{P}(N)$ containing all the $\sigma^{z}$ Pauli strings. The fact that the Pauli strings that count are those belonging to $\mathbb{Z}$ can be also understood by looking to the Hamiltonian in Eq.~\eqref{isinghamiltonian}: For $\lambda\gg1$ the dominant term is $\lambda\sum_{i}\sigma_{i}^{z}$ whose eigenstates are stabilizer states belonging to the stabilizer group $\mathbb{Z}$. In other words, we are estimating the average in Eq.~\eqref{magicintermsofcorrelations} by (importance) sampling the probability distribution with Pauli strings $P\in\mathbb{Z}$. Thus, the estimated density of non-stabilizerness can be computed as 
\be
\alpha(\lambda)\simeq-\frac{1}{N}\log_2\frac{\sum_{\{i\}_{k},\{j\}_k\le N}G_{0}(\lambda)^{4}\delta_{\{i\}_k}^{\{j\}_k}}{\sum_{\{i\}_{k},\{j\}_k\le N}G_{0}(\lambda)^2\delta_{\{i\}_k}^{\{j\}_k}}
\label{importancesampling}
\ee
where we introduced a normalization over the sampling given by $\sum_{\{i\}_{k},\{j\}_{k}\le N}\mathcal{C}(\{i\}_k,\{j\}_k)^{2}$, cfr. Eqs.~\eqref{magicdefinition} and \eqref{magicintermsofcorrelations}. The straightforward computation of Eq.~\eqref{importancesampling}, together with the fact that $G_{0}(\lambda)^2=\aver{\sigma^{z}}^2$ leads to Eq.~\eqref{densitymatrix1sitemagic}. Thus, the density of non-stabilizerness estimated by importance sampling does coincide with the $L-$density of non-stabilizerness with $L=1$.

The fact that one can access non-stabilizerness from local measurements is nontrivial and in general, is not true. We can show it by considering a simpler example: Suppose having a bipartite system $AB$, a random pure state $\ket{\Psi_{AB}}$ and consider the percent different in non-stabilizerness $\epsilon_{AB}=(M_{AB}-M_{A}-M_{B})/M_{AB}$; here $M_{AB},M_{A},M_{B}$ are the stabilizer R\'enyi entropies of $\ket{\Psi_{AB}}$ and $\rho_{A}=\tr_{B}(\ket{\Psi_{AB}}\bra{\Psi_{AB}})$ and $\rho_{B}$ respectively. Thanks to the typicality of the stabilizer R\'enyi entropy\cite{leone2021renyi} and the two-R\'enyi entropy\cite{popescu2006entanglement} over the set of Haar-random states, one gets $\epsilon_{AB}\approx 1$ (up to an exponentially small error in $\dim(AB)$), which means that the non-stabilizerness cannot be accessed locally for the majority of states in the Hilbert space. The above argument can be straightforwardly generalized to the case of the multipartite system $A_{1}A_{2}\cdots A_{h}$.

{\em Conclusions and Outlook.---} The complex pattern of the ground-state wave-function of a quantum many-body system depends on the interplay between its entanglement and the non-Clifford resources, or non-stabilizerness, that it contains. Although both in the gapped phase and at the critical point the ground state of the transverse field Ising model contains an extensive amount of non-stabilizerness, away from criticality this is localized. On the other hand, at the critical point, its non-stabilizerness is delocalized and described by a power law. 

These results raise a number of questions. First, one could extend these methods to models featuring localization through disorder or frustration. One expects that any form of localization would result in being able to evaluate non-stabilizerness by few-site quantities. Second, the same methods can be used to study the dynamics of a quantum many-body system after a quench. It would be interesting to see whether non-stabilizerness delocalizes as the system evolves in time and if equilibration ensues. Moreover, it is very intriguing to study the behavior of non-stabilizerness in such systems when integrability is broken. The role of quantum complexity implied in the conjunction of non-stabilizerness and entanglement for the onset of thermalization and non-integrable behavior has been recently studied in the context of doped quantum circuits\cite{leone2021quantum,oliviero2021transitions,true2022transitions} and Hamiltonians\cite{leone2020isospectral,Oliviero2020random}, but a local quantum many-body system is its most natural setting. The main result of this paper opens the way to the experimental measurement of non-stabilizerness by local measurements, for instance, in ultra-cold atom gases realizing the Bose-Hubbard model. Finally, although further investigation is necessary, we can argue that the delocalization of non-stabilizerness at the critical point suggests that the CFT theory, underlying critical many body systems, enjoys delocalization of non-stabilizerness as well.

{\em Acknowledgments.---} The authors thank Francesco Caravelli, Stefano Piemontese and Seth Lloyd for inspiring discussions and comments. The authors acknowledge support from NSF award no. 2014000. The work of L.L. and S.F.E.O. was supported in part by College of Science and Mathematics Dean’s Doctoral Research Fellowship through fellowship support from Oracle, project ID R20000000025727.

\end{document}